\documentclass[%
reprint,
amsmath,amssymb,
aps,
pra,
]{revtex4-1}
\usepackage{graphicx}
\usepackage{CJK}   

\begin{document}

\title {
Molecular near-infrared transitions determined with sub-kHz accuracy
    }
	
\author{
Jin Wang$^{1}$, Yu R. Sun$^{1,2}$, Lei-Gang Tao$^{1}$, An-Wen Liu$^{1,2}$, Shui-Ming Hu$^{1,2}$}
\email{Corresponding author. smhu@ustc.edu.cn}

\affiliation {
$^1$ Hefei National Laboratory for Physical Sciences at Microscale, $i$Chem center,
        University of Science and Technology of China, Hefei, 230026 China; \\
$^2$ CAS Center for Excellence and Synergetic Innovation Center
 in Quantum Information and Quantum Physics,
        University of Science and Technology of China, Hefei, 230026 China}

\begin{abstract}
Precise molecular transition frequencies are needed
 in various studies including the test of fundamental physics.
Two well isolated ro-vibrational transitions
 of $^{12}$C$^{16}$O at 1.57~$\mu$m,
 R(9) and R(10) in the second overtone band,
 were measured by a comb-locked cavity ring-down spectrometer.
Despite the weakness of the lines (Einstein coefficient $A\simeq 0.008$~s$^{-1}$),
 Lamb-dip spectra were recorded with a signal-to-noise ratio over 1000,
  and the line positions were determined to be
  191 360 212 761.1 and 191 440 612 662.2~kHz respectively,
  with an uncertainty of 0.5~kHz ($\delta\nu/\nu=2.6\times 10^{-12}$).
The present work demonstrates the possibility to explore extensive molecular lines
 in the near-infrared with sub-kHz accuracy.
\end{abstract}



\maketitle

Rotation-resolved molecular transitions are convenient natural frequency grids
 used in spectroscopy,
  communication~\cite{Hong2003OL_C2H2},
  metrology~\cite{Foreman2005OL_CH4_Clock}
  and astrophysics~\cite{Smith2009ApJ_CO}.
Their precise positions are also concerned in fundamental physics,
 being used to determine physical constants,
  either the values~\cite{Biesheuvel2016NC_HD}
  or the space-time variation~\cite{Hudson2006PRL_OH,Shelkovnikov2008PRL_SF6_mu,Ubachs2016RMP_H2},
  to find parity-violation effects~\cite{Daussy1999PRL_CHFClBr},
  and even to search new physics beyond the Standard Model~\cite{Salumbides2013PRD}.
Accuracy is crucial in these measurements.
For example, a fractional uncertainty of $10^{-15}$ to $10^{-20}$
 is needed~\cite{Quack2008ARPC_Chiral,Tokunaga2013MP_Chiral}
 to detect parity-violation effects in chiral molecules.

Absorption spectroscopy is the most frequently employed method
 to derive molecular line parameters used in miscellaneous applications,
 but often lacks precision and sensitivity.
By putting sample gases in an optical resonant cavity of high finesse,
 the sensitivity can be enhanced by orders of magnitude to detect extremely weak transitions~\cite{Kassi2012JCP_CRDS,Tan2014JMS_H2}.
Meanwhile, the resonance effect also considerably enhances the intra-cavity light intensity,
 which allows saturation spectroscopy measurements using low-power continuous-wave
  lasers~\cite{Giusfredi2010PRL_CRDS_CO2_Sat,Gatti2016SR_Comb_Sat,Wang2017RSI}.
In recent years, considerable efforts~\cite{Gatti2016SR_Comb_Sat, Wang2017RSI, Truong2013JCP_Comb_CRDS, Santamaria2016PCCP_Comb_CRDS_BuffCooling, Spaun2016Nat_Comb_LowT, Burkart2015JCP_CRDS_Sat}
 have been carried out to pursue both high sensitivity and high precision
 by combining resonant cavities and frequency combs.
The method has been applied to determine absolute infra-red line positions
 from Lamb-dip measurements of several molecules~\cite{Gambetta2010NJP_H2O_17_18, Hong2003OL_C2H2,Balling2005OE_C2H2,Edwards2005APB_C2H2_13, Wang2017RSI, Madej2006JOSAB_C2H2_13,Twagirayezu2015JMS_C2H2, Gatti2016SR_Comb_Sat,Czajkowski2009OE_NH3, Burkart2015JCP_CRDS_Sat,Giusfredi2010PRL_CRDS_CO2_Sat,Amy2004JMS_CO2_9um,
 Okubo2011OE_CH4_3u, Ting2014JOSAB_N2O_4u5}.
Most of the reported molecular line positions have an uncertainty at the kHz level
 (fractional accuracy $10^{-11}$),
 including the C$_2$H$_2$ line positions near 1.54~$\mu$m recommended
 by Bureau International des Poids et Mesures
 for practical realization of the metre and secondary representations of the second.

Here we present a method that enables Lamb-dip absorption spectroscopy
 of weak molecular transitions with
 a frequency accuracy of sub-kHz and a sensitivity of $10^{-12}$~cm$^{-1}$ as well.
The positions of two ro-vibrational transitions of $^{12}$C$^{16}$O at 1.57~$\mu$m,
 with Einstein coefficients of about 8~mHz,
 were determined with an accuracy of 0.5~kHz ($\delta\nu/\nu=2.6\times 10^{-12}$).
To the best of our knowledge, it is the first time that
 near infrared (NIR) transitions of molecules
 have been determined with sub-kHz accuracy.
The CO molecule was chosen here because
 it is one of the most extensively studied and prototype diatomic molecule.
As the second most abundant neutral molecule in the universe
 only after the molecular hydrogen,
 CO plays a key role in the gas-phase chemistry in the interstellar medium.
The spectroscopic parameters of CO are also of great needs in
 astrophysical studies~\cite{Li2015ApjSS_CO}.
These two well-isolated transitions, R(9) and R(10) lines in the 3-0 band,
 were also selected~\cite{Cheng2015Met_kB}
  for the optical determination of the Boltzmann constant~\cite{Daussy2007PRL_Boltzmann,Sun2011OE},
 where both precision and sensitivity are critical.
The method can be applied in various studies,
 for example, to improve the frequency standards in the near-infrared,
 and to determine the proton-electron mass ratio
 from the ro-vibrational spectroscopy of the HD molecule.

\begin{figure}[htp]
	\centering
	\includegraphics[width=3.5in]{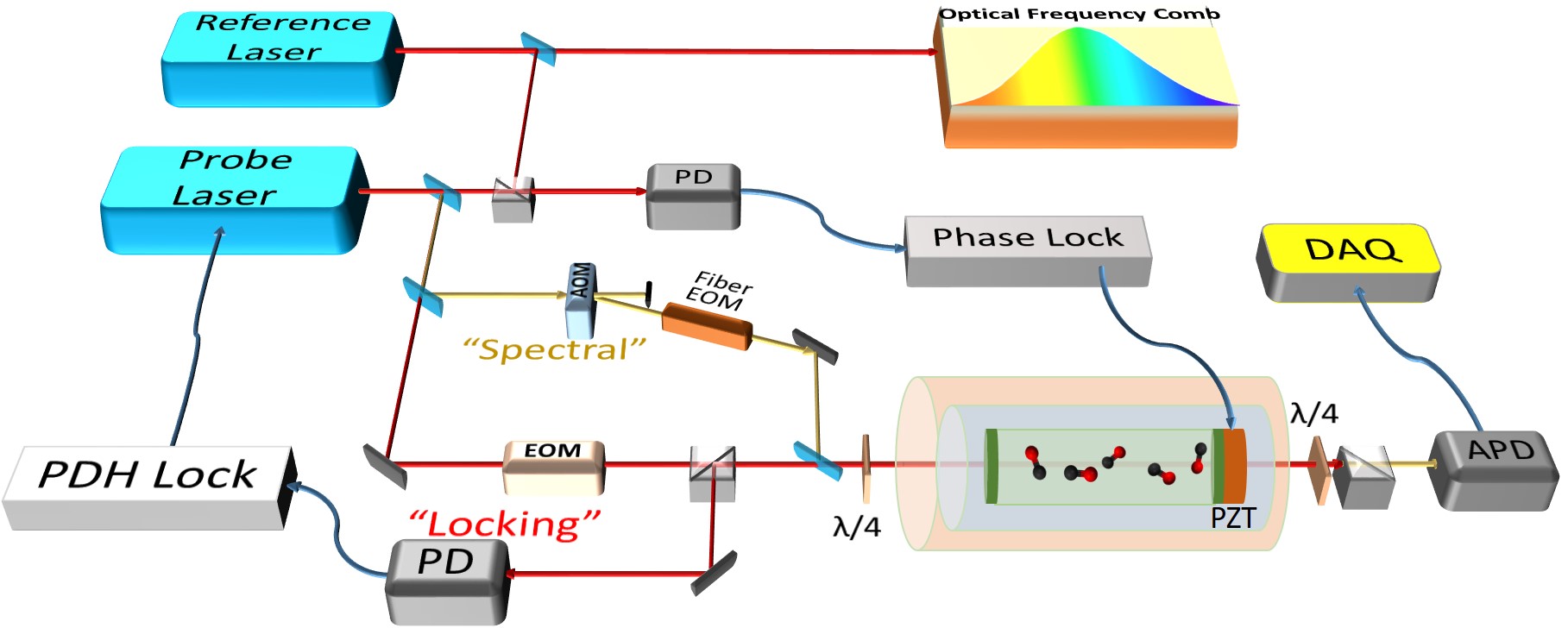}
	\caption{ Configuration of the experimental setup.
        The probe laser frequency is locked with the cavity
         (the ``locking'' beam).
        Another beam from the probe laser (referred to as ``spectral'' beam)
         is frequency shifted and used for CRDS measurement.
        The ring-down cavity length is locked according to the beat signal between
         the probe and reference lasers.
		Abbreviations:
		AOM: acousto-optical modulator;
		APD:  avalanche photodiode detector;
        DAQ: data acquisition system;
		EOM: electro-optical modulator;
        PD:  photodiode detector;
		PZT: piezo actuator.
		\label{Fig_Setup}
	}
\end{figure}
\begin{figure}[htp]
	\centering
	\includegraphics[width=3.5in]{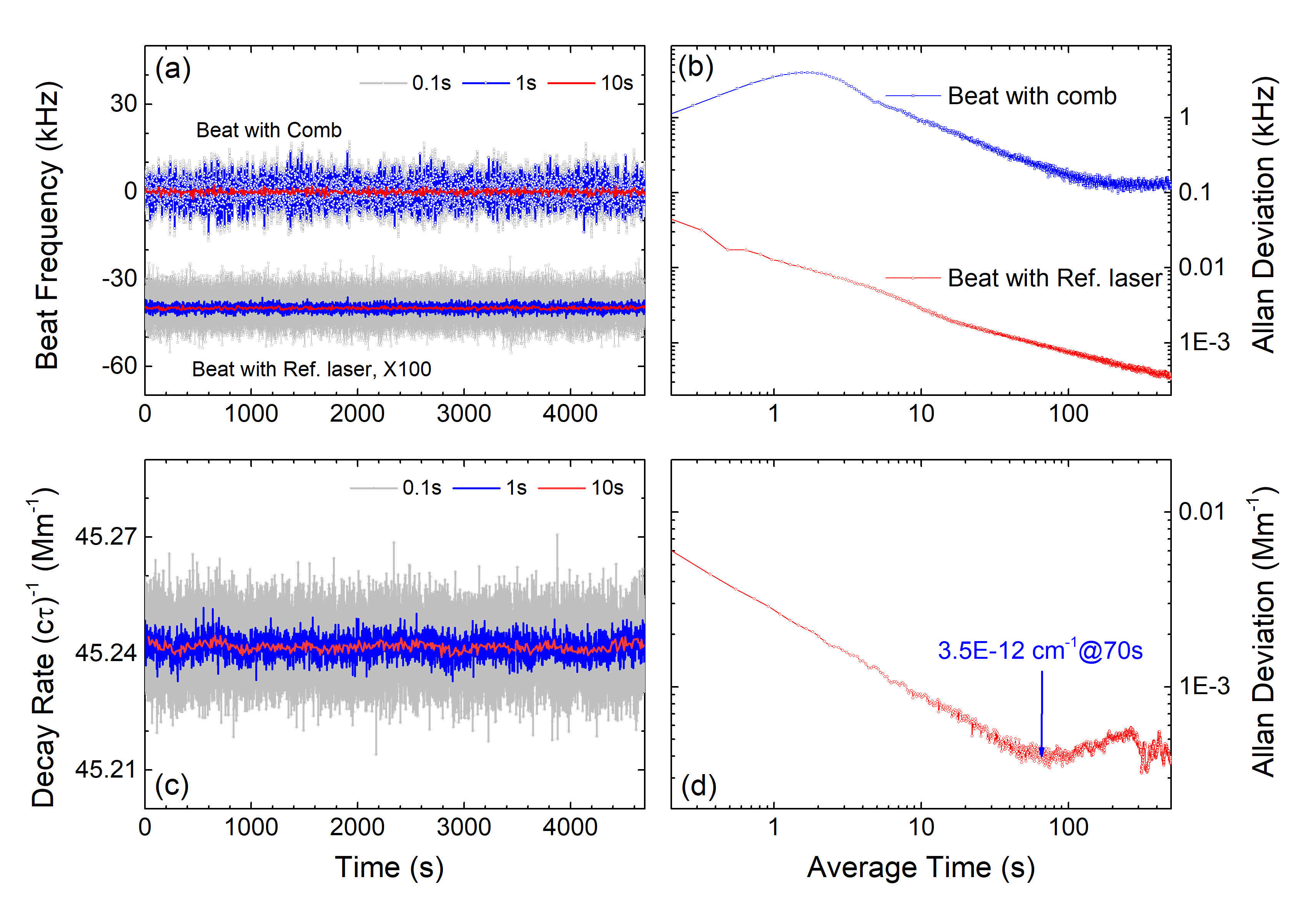}
	\caption{
		The frequency precision and sensitivity of the CRDS instrument.
		(a) Beat frequencies among the probe laser, the reference laser and the frequency comb.
		Data are shifted for better illustration.
		(b) Allan deviation of the beat frequencies.
		(c) Absorption coefficient detected with an empty cavity.
		(d) Allan deviation of the data shown in (c).
		\label{Fig_Allan}
	}
\end{figure}

The configuration of the experimental setup is presented in Fig.~\ref{Fig_Setup}.
The probe laser (Koheras NKT E15) is locked to
 a ring-down (RD) cavity using the Pound-Drever-Hall (PDH) method.
The RD cavity has a finesse of $1.56\times 10^5$
 and a free-spectral-range (FSR) of 336~MHz.
The 50~cm-long RD cavity is installed in a stainless-steel vacuum chamber
 and its temperature is stabilized at 0$^\circ$C
 with a fluctuation below 1~mK,
 detected by two calibrated thermal sensors attached on two sides of the cavity.
Owing to the thermo-stability of the cavity,
 the day-to-day drift of the FSR value is below 10~Hz,
 much less than the linewidth of the cavity mode (2.2~kHz).
The cavity length is stabilized through a piezo actuator (PZT)
 driven by a phase-lock circuit based on
 the beat signal between the probe laser and a reference laser.
The reference laser is locked to
 a Fabry-P\'{e}rot interferometer made of ultra-low-expansion glass,~\cite{Gao2010RSI_CRDS}
 and its absolute frequency is calibrated by
 a frequency comb which is synthesized by a Er:fiber oscillator
 operated at 1.56~$\mu$m.
The repetition frequency ($f_R\approx 180$~MHz)
 and carrier offset frequency ($f_0$) of the comb
 are locked to precise radio-frequency sources,
 both referenced to a GPS-disciplined rubidium clock (SRS FS725).
The beat signals among the probe laser, the reference laser, and the comb
 were recorded by a frequency counter (Agilent 53181A),
 and they are illustrated in Fig.~\ref{Fig_Allan}.
We can see that the probe laser frequency follows well with the reference laser.
The fast fluctuation of the beat signal between the probe laser and the comb
 comes mainly from the line width of the comb tooth,
 while the long-term drift is due to
 the slow drift (2~kHz/hr) of the reference laser.

A separated beam from the probe laser,
 which passes an acousto-optic modulator (AOM)
  and a fiber electro-optic modulator (EOM),
 is used to produce the ring-down signal.
The input laser power to the fiber EOM is about 3~mW,
 and about 1/3 of the power is delivered to one of the 1st-order sidebands.
The laser light transmitted from the cavity is
 detected by an avalanche photodiode detector (APD).
The transmittance of an empty cavity is about 3~\%,
 leading to~\cite{Ma1999JOSAB_C2H2,Wang2017RSI}
  an intra-cavity laser power of about 8~W.
A combination of polarizing waveplates and Glan-Tylor prisms was used
 to separate the two beams for frequency locking and spectral probing.
In addition, because of the significant frequency difference,
 the beat signal between the two beams is out of the bandwidth of the detectors.
Finally, the interference between the frequency locking and RD detection signals
  has been reduced to a negligible level.
The frequency $f_{AOM}$ used to drive AOM is fixed at 80.0009~MHz
 and the frequency $f_{EOM}$ used to drive EOM is tuned to
  fulfill the equation: $f_{AOM}+f_{EOM}= f_{M+n}-f_M$,
  where $f_M$ is the frequency of the cavity mode which the probe laser is locked with,
   $f_{M+n}$ is the frequency of a nearby cavity mode with $n$ value of about 3.
In this case, only one 1st-order sideband in the beam
 is on resonance and passes the cavity,
 while the carrier and other sidebands are completely reflected.
In this way, the frequency of the laser beam for CRDS detecting is set to be:
 $\nu = \nu_{ref} + f_B + f_{EOM} + f_{AOM}$.

The acousto-optic modulator is also used to periodically chop the spectral beam
 to initiate ring-down events.
The ring-down curve is fit by an exponential decay function
 to derive the decay time $\tau$,
 and the sample absorption coefficient $\alpha$ is determined
  according to the equation: 
  $\alpha = (c\tau)^{-1} - (c\tau_0)^{-1}$,
 where $c$ is the speed of light,
  and $\tau_0$ is the decay time of an empty cavity.
Fig.~\ref{Fig_Allan}(c) shows the recorded $(c\tau)^{-1}$ values of the empty cavity,
 and the corresponding Allan deviation is given in Fig.~\ref{Fig_Allan}(d).
The limit of detection, presented as the minimum detectable absorption coefficient,
 reaches about $3.5\times 10^{-12}$~cm$^{-1}$ at an averaging time of about 70 seconds,
 being comparable to those best records obtained by cavity enhanced absorption spectroscopy~\cite{Truong2013NatPhoton_FARS_CRDS,Foltynowicz2011PRL_FTComb,Burkart2014OL_OFFS_CRDS}.
Compared to our previous method~\cite{Wang2017RSI},
 the frequency lock is kept
 when we chop the spectral beam for ring-down measurements,
 resulting with an improved frequency lock.
Precise temperature control of the cavity is also essential
 to maintain a satisfactory frequency stability during the scan.
Moreover, the present method allows us to synchronize
 ring-down events to an external trigger source.
It is very useful when the sample molecule is short-lived species,
 such as transient species produced in photolysis.
When a stable gas samples is used,
 an advantage is to maximize the repetition rate of the RD events,
 which is only limited by the decay and build-up time of each ring-down curve.
In this study, the repetition rate is about 1.5~kHz,
 over one order of magnitude higher than that obtained in our previous study,
 leading to a considerably improved sensitivity.

\begin{figure}[htp]
	\centering
	\includegraphics[width=3.5in]{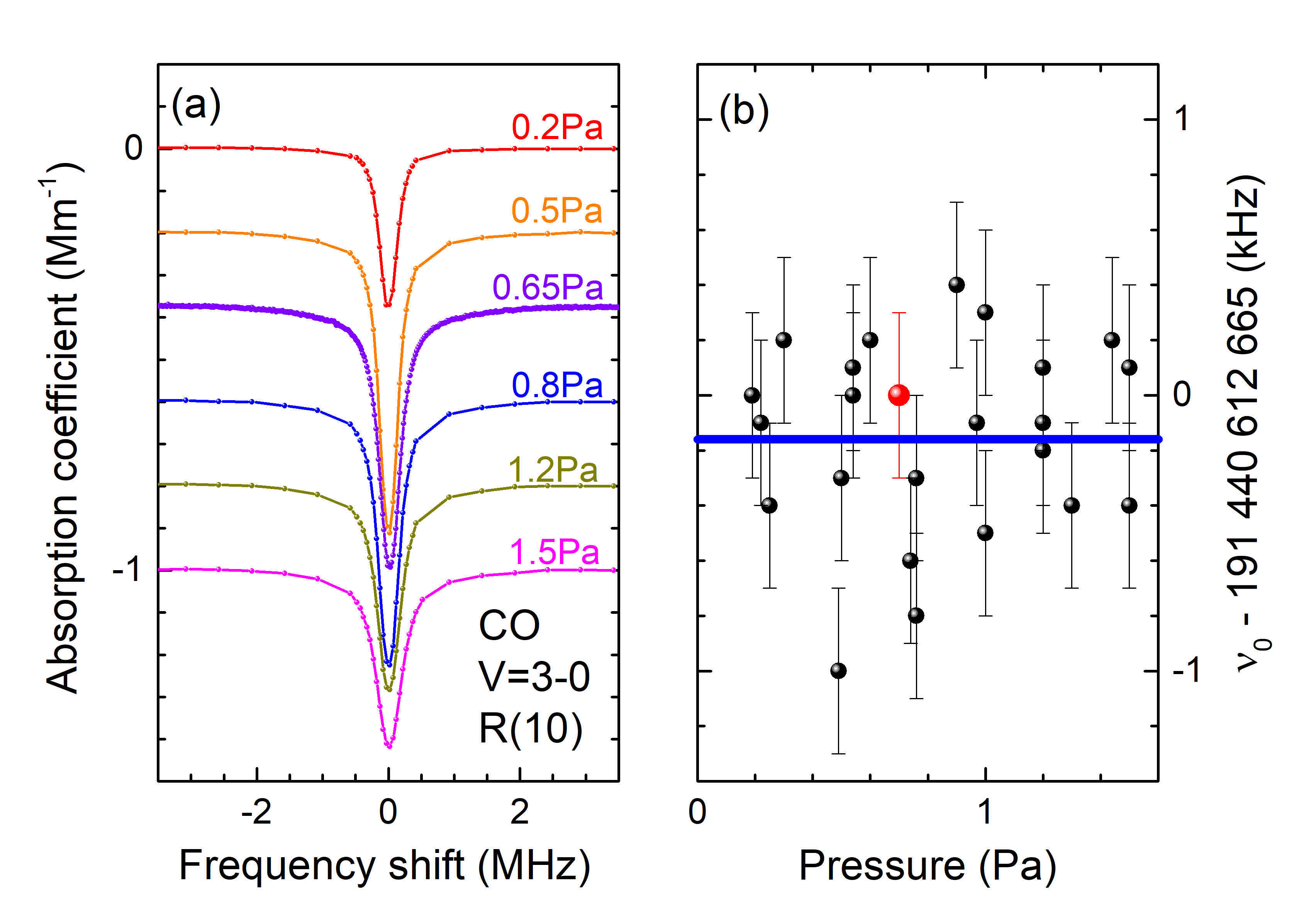}
	\caption{
    (a) Cavity ring-down saturation spectra of the R(10) line
        in the (3-0) band of $^{12}$C$^{16}$O.
	(b) positions of the R(10) line determined from
        spectra recorded with different sample pressures.
        The red point indicates a measurement took about half year later than
         other measurements (black points).
		The blue line indicates the weighted average of the data.
		\label{Fig_R10}
	}
\end{figure}

Natural carbon monoxide sample gas was used in the measurement.
As given in the HITRAN database~\cite{HITRAN2012},
 R(9) and R(10) lines in the $V=3-0$ overtone band of $^{12}$C$^{16}$O
 are located at 6383.09~cm$^{-1}$ and 6385.77~cm$^{-1}$, with intensities of
 $2.03\times 10^{-23}$cm/molecule and $1.88\times 10^{-23}$cm/molecule, respectively.
The saturation power is estimated to be about 1.5~kW~cm$^{-2}$,
 according to the formulas given by
 Giusfredi \textit{et al.}~\cite{Giusfredi2010PRL_CRDS_CO2_Sat}
Note that the intra-cavity power also decays during a RD event,
 which reduces the saturation effect
 and leads to an non-exponential decay~\cite{Giusfredi2010PRL_CRDS_CO2_Sat}.
Since the main purpose of the present study is to determine the precise line position
 instead of the profile of the saturation spectrum,
 we simply applied a single exponential decay function to fit the RD curve.
Only data within the beginning 50~$\mu$s of each RD curve were included in the fit,
 while the ring-down time was about 30~$\mu$s at a sample pressure of 1~Pa.
The intra-cavity light intensity was estimated to be from 3~W to 0.5~W
 during this period.
Taking into account a laser beam waist radius of 0.76~mm,
 we estimated that the saturation parameter was below 10\%.
The spectral scan was accomplished by tuning the frequency $f_B$.
At each frequency point, about 3000 RD events were recorded within 2 seconds.
Each scan took about 100 seconds
  when a typical frequency step of 50~kHz around the line center was applied.
We also tried with finer frequency steps but did not see notable differences.
Spectra were recorded with sample pressures in the range between 0.2~Pa and 1.5~Pa,
 and some of them are shown in Fig.~\ref{Fig_R10}(a).
Each spectrum shown in the figure is an average of 20 scans
 and the signal-to-noise ratio reaches over 1000.

The line parameters, position, depth, and width, are derived by fitting the spectra.
For simplicity, the Lorentzian function was used in the fit,
 although deviation from the Lorentzian profile has been found in
 saturation spectra~\cite{Borde1976PRA_SaturationProfile,Chardonnet1994APB_OsO4}.
The derived line width (full width at half maximum) is about 0.2~MHz,
 being consistent with the transit-time broadening width of 0.13~MHz
 estimated from the mean velocity of the CO molecules
 and the laser beam waist.
Fig.~\ref{Fig_R10}(b) shows the R(10) line positions obtained at different pressures,
 and each data point stands for an averaged value derived from about 30 scans.
The red point in Fig.~\ref{Fig_R10}(b) is the result from the measurement
 taken half a year after other measurements (black points).
The excellent agreement indicates the long-term stability of the measurements.
The horizontal line shown in Fig.~\ref{Fig_R10}(b) gives the weighted average of the
 line positions obtained at different pressures
 and the statistical uncertainty is 0.1~kHz.
Other contributions to the uncertainty have been investigated, described below.

If we use a linear fit to the data shown in Fig.~\ref{Fig_R10}(b),
 we can get a coefficient of $0.2\pm 0.2$~kHz/Pa.
However, a self-pressure induced shift of -189~MHz/atm (1.9~kHz/Pa) has been
 reported~\cite{Picque1997JMS_CO_30} based on the Doppler-broadened spectra of CO
 recorded under normal pressures ($\sim 10^4$~Pa).
There is a considerable difference between
  the line shift obtained from saturation spectra
 and that from conventional Doppler-limited spectra.
Similar results~\cite{Borde1976PRA_SaturationProfile, Chardonnet1994APB_OsO4, Ma1999JOSAB_C2H2}
  have been reported before, but it has not yet been well interpreted.
Here we simply use the value averaged from the positions
 obtained from spectra recorded at different pressures,
 but an uncertainty of 0.2~kHz is left for possible contribution from
 the pressure-induced shift which needs further investigation.
We also include an uncertainty of 0.2~kHz due to the model of the line profile,
 although we did not observe any asymmetry above the noise level.

The spectral laser frequency is determined by
 the reference laser frequency $\nu_{ref}$, the beat frequency $f_B$,
 and the radio frequencies applied on AOM and EOM.
The reference laser frequency is calibrated by the frequency comb
 and eventually by the GPS-disciplined rubidium clock.
The rubidium clock has a stated frequency stability
 of $1\times 10^{-11}$ at 10~s and
 the absolute frequency accuracy is $2\times 10^{-12}$ (0.4~kHz at 1.57~$\mu$m)
 when disciplined by the GPS signal.
The beat frequency $f_B$ is monitored by a frequency counter
 which shows a fluctuation well below 50~Hz (Fig.~\ref{Fig_Allan}).
The RF frequency $f_{AOM}$ applied on the AOM has a fluctuation below 50~Hz,
 mainly due to the drift of the room temperature.
The EOM driven frequency $f_{EOM}$ is also referenced to the rubidium clock,
 and the accuracy is better than 1~Hz.
Taking a root-square mean velocity of 493~m/s of the CO molecule at 273~K,
 we estimate that the shift due to the second order Doppler effect is -258~Hz
  and the related uncertainty is below 10~Hz.
Under a laser power of 3~W inside the cavity,
 the AC Stark shift is negligible for these two CO transitions.
We have also used a different input laser power in the measurement,
 but found no difference in the line position within the experimental uncertainty.

The overall uncertainty budget is given in Table~\ref{TableBudget}.
The positions for lines R(9) and R(10) are determined to be
 191 360 212 761.1~kHz and 191 440 612 662.2~kHz respectively,
 with an overall uncertainty of 0.5~kHz.
The R(9) line position agrees with the value of 191 306 212 770(7)~kHz
 from our previous study~\cite{Wang2017RSI},
 but the accuracy has been improved by an order of magnitude.
Note that a recoil shift of -2.9~kHz has been included in this work
 but not in our previous study.
The results also agree with the line positions reported by
 Mondelain \textit{et al}~\cite{Mondelain2015JQSRT_CO_30_CRDS}
 from Doppler-limited CRDS study,
 which were 191 360 212.54~MHz and 191 440 612.43~MHz
  with an uncertainty of 0.3~MHz.

\begin{table}[htp]
\centering
\caption{Uncertainty budget, R(9) and R(10) lines in the $V=3-0$ band of $^{12}$C$^{16}$O
(unit: kHz).
\label{TableBudget}
}
\begin{tabular}{|lll|}
\hline
Source & \multicolumn{1}{c}{Frequency} & Uncertainty \\
\hline
Statistical & 191 360 212 763.7 (R9)  & 0.1 \\
            & 191 440 612 664.8 (R10) & 0.1 \\
Comb frequency & & 0.4   \\
Locking servo  & & 0.05 \\
EOM frequency  & & 0.001\\
AOM frequency  & & 0.05    \\
Pressure shift & & 0.2  \\
Line profile asymmetry & & 0.2 \\
2nd order Doppler & +0.26 & 0.01 \\
Recoil shift   & -2.9 & - \\
\hline
Total &191 360 212 761.1 (R9)& 0.5 \\
      &191 440 612 662.2 (R10)& 0.5 \\
\hline
\end{tabular}
\end{table}


In conclusion, a laser-locked cavity ring-down spectrometer,
 with both the probe laser and the high-finesse cavity locked
  and referenced to an optical frequency comb,
 was constructed and used to record
 the Lamb-dip spectra of the R(9) and R(10) lines in the $V=3-0$ band of CO.
The present work demonstrates an accuracy of 0.5~kHz
 and also a sensitivity of $3.5\times 10^{-12}$~cm$^{-1}$.
Note that the accuracy is currently limited by the rubidium clock,
 the line profile model, and the collision effect,
 while the statistical uncertainty is only 0.1~kHz.
By using a more accurate clock (like a hydrogen maser or a fountain clock),
 and taking further analysis on the Lamb-dip line profile,
 we expect to pursue an accuracy of 0.1~kHz or better.
There are an enormous amount of molecular transitions in the near infrared
 with a transition rate higher than the CO lines studied here,
 and their positions can be determined with the same method,
 which could considerably enlarge and improve the near-infrared frequency standards.
We are also using the method to detect the saturation spectrum of the HD molecule.
The strongest absorption line of HD accessible by a NIR diode laser
 is the R(1) line in the first overtone, located at 7241~cm$^{-1}$,
 with a saturation power of $3\times 10^7$~W~cm$^{-2}$.
Using an external-cavity diode laser with several tens of milli-Watt,
 we would be able to detect the Lamb-dip at a sensitivity of $10^{-12}$~cm$^{-1}$
 and expect to determine the line position with a fractional accuracy of $10^{-10}$.
Combined with the progress on the theoretical calculation~\cite{Puchalski2016PRL_H2}
 of this two-electron molecule,
 such a measurement will provide an independent determination of
 the proton-electron mass ratio.

The authors are indebted to Dr. P. Wcis{\l}o from Nicolaus Copernicus University
for helpful discussions.
This work is jointly supported by CAS (XDB21020100), by NSFC (21688102, 91436209, 21427804, 21225314), and by NBRPC (2013CB834602).

%

\end{document}